# Neutronic performances of the MEGAPIE target

S. Panebianco[1,a], P. Beauvais[2], O. Bringer[1], S. Chabod[1], F. Chartier[3], E. Dupont[1], A. Letourneau[1], P. Lotrus[2], L. Oriol[4], F. Molinié[2], and J. Ch. Toussaint[2]

[1] CEA Saclay, DSM/DAPNIA/SPhN, F-91191 Gif-sur-Yvette Cedex, France
[2] CEA Saclay, DSM/DAPNIA/SIS, F-91191 Gif-sur-Yvette Cedex, France
[3] CEA Saclay, DEN/DPC/SECR, F-91191 Gif-sur-Yvette Cedex, France
[4] CEA Cadarache, DEN/DER/SPEX, F-13108 Saint-Paul-lez-Durance, France

**Abstract.** The MEGAPIE project is a key experiment on the road to Accelerator Driven Systems and it provides the scientific community with unique data on the behavior of a liquid lead-bismuth spallation target under realistic and long term irradiation conditions. The neutronic of such target is of course of prime importance when considering its final destination as an intense neutron source. This is the motivation to characterize the inside neutron flux of the target in operation. A complex detector, made of 8 "micro" fission-chambers, has been built and installed in the core of the target, few tens of centimeters from the proton/Pb-Bi interaction zone. This detector is designed to measure the absolute neutron flux inside the target, to give its spatial distribution and to correlate its temporal variations with the beam intensity. Moreover, integral information on the neutron energy distribution as a function of the position along the beam axis could be extracted, giving integral constraints on the neutron production models implemented in transport codes such as MCNPX.

## 1 Introduction

Based on an initiative of six European institutions (PSI, FZK, CEA, SCK-CEN, ENEA, CNRS), JAEA (Japan), DOE (US) and KAERI (Korea), the MEGAwatt Pilot Experiment (MEGAPIE) started officially in the year 2000 aiming to design, build and safely operate a liquid metal (Lead-Bismuth Eutectic, LBE) spallation target at 1 MW beam power [1]. It is considered as an essential step towards the development of high power spallation targets to produce intense neutron sources or neutrino beams and a fundamental technological piece for nuclear waste incinerators driven by accelerators (ADS). In particular, it is meant to be the key experiment for high power window targets within the transmutation activities. The experience gained in heavy liquid metal technology will also contribute to the lead-cooled reactor research within the GEN-IV initiative. The target was delivered to PSI in 2005, installed and tested in the SINQ hall during spring 2006 and successfully irradiated during 4 months starting in mid-August.

Although spallation models are nowadays reliable and well qualified against a lot of experimental nuclear data, a spallation target as the MEGAPIE one is a complex system for which the accuracy on the neutron flux simulation is still questionable. Moreover, a precise neutronic characterisation is crucial for future ADS developments and to address the possibility to transmute minor actinides in such a system. These are the reasons why we have designed and built a neutron detector to measure "in situ" and to characterise the inner neutron flux of the target under irradiation. Coupled with very detailed Monte Carlo simulations, these integral measurements should provide the scientific community with accurate data on the neutron generation of such a system to constraint commonly used neutron transport codes and neutron production models. Moreover, other effects as the influence of spallation residues accumulation on the neutron balance or the temperature on the neutron yield could be assessed. Finally, the incineration of two important actinides should be studied in a neutron flux representative of a moderated spallation source and of what could provide the next generation of future systems (High Temperature Reactor or Molten Salt Reactor).

## 2 Description of the experiment

### 2.1 The MEGAPIE target

The target, installed in the SINQ location at the Paul Sherrer Institute (Switzerland), has been designed to accept a proton current of 1.74 mA. The thermal energy deposited in the lower part of the target is removed by forced convection. The LBE is driven by the main inline electromagnetic pump, then passes through a 12-pin heat exchanger (HX) and returns to the spallation region. The heat is evacuated from the heat exchanger through a diathermic oil loop to an external intermediate water cooling loop and then finally goes into the PSI existing cooling system. The beam entrance window is cooled both by the main flow and also by a cold LBE jet extracted at the heat exchanger outlet, which is pumped by a second electro magnetic pump. The target has been conceived in nine sub-components, which were manufactured separately and finally assembled.

---

[a] Presenting author, e-mail: stefano.panebianco@cea.fr

### 2.2 Neutron Detector generalities

We have designed and developed an innovative detector to measure the absolute inner neutron flux of the target with accuracy better than 5% and to follow the time and spatial evolutions of the flux over a large neutron intensity range. It should be noticed that a 1 MW liquid Pb-Bi spallation target like MEGAPIE constitutes a very constraining environment due to:

- *high temperature*: around 420 °C with beam-on and 230 °C with beam-off,
- *high level of radiations*: more than $10^{13}$ n/cm$^2$/s and almost the same level of gamma rays coming from spallation reactions and activation of structural materials,
- *electromagnetic perturbations* due to electromagnetic pumps.

Moreover, the overall dimensions of the central rod where the detector was located were very tiny (20 mm in diameter) and its access impossible during the whole irradiation period.

The neutron detector, built in 2005, contains eight micrometric fission-chambers already employed in the framework of the Mini-INCA project [3]. The fission chambers were adapted for the MEGAPIE specific environment, with dedicated cables, electronics and acquisition system [4]. The detector is 5 meters long with a 13 mm diameter in its lower part and 22 mm in the upper part. Fission chambers are located in the thinner part of the detector, to be as closed as possible from the proton-beam interaction zone in the Pb-Bi (fig. 1).

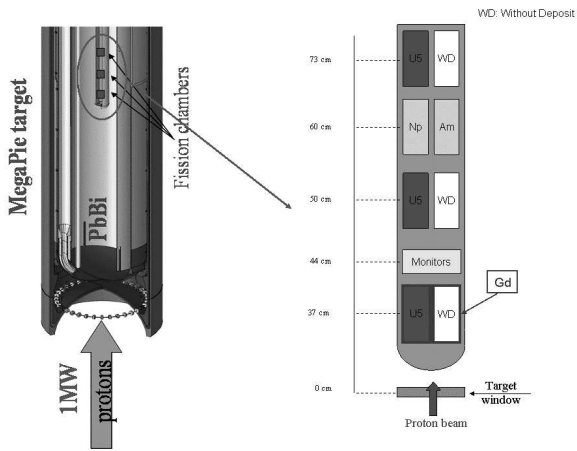

**Fig. 1.** Schematic view of the bottom part of the MEGAPIE target (left) and neutron detector layout (right).

Signals from fission chambers are transported by 1 mm mineral cables inside the detector and connected to triaxial organic cables outside the detector to avoid electromagnetic perturbations. Fission chambers are imbedded in pairs along the axis of the detector over a 50 cm length. Each pair, except one, is made by a chamber containing $^{235}$U fissile isotope and a chamber without deposit to compensate the fission signal from leakage currents or from currents induced by radiation fields. Cables were chosen to prevent leakage current higher than few nA at 500°C temperature. The bottom pair is shielded with natural Gd filter to absorb thermal neutrons and be more sensitive to epithermal neutrons. Finally, one pair is constituted by a chamber with $^{241}$Am and the other one with $^{237}$Np to probe their incineration. These different configurations are chosen to provide an overall characterisation of the inside neutron flux, in terms of its intensity but also its energy distribution. To increase the accuracy on the energy spectrum determination, nine activation neutron flux monitors were put inside the detector in a small Ti-box. These monitors will be analysed during the post-irradiation phase (2008-2009). A schematic view of the neutron detector layout and the simulated neutron spectra are shown respectively on fig. 1 and fig. 2.

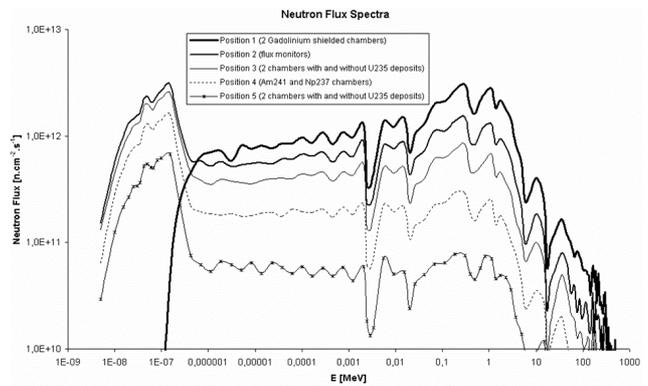

**Fig. 2.** Simulated neutron fluxes corresponding to the different positions along the beam axis. Positions 1 and 5 correspond to the lower and upper chambers respectively.

### 2.3 Systematic errors reduction

The reduction of systematic errors is one of the challenges for these measurements. Indeed, fission chambers are usually used in relative measurements where the sensitivity of the detector is an effective observable calibrated with respect to a reference. In our case, relative calibration was not feasible in absence of the neutron flux. Thus, a deep validation and calibration campaign of all the detector pieces, the electronics and the acquisition, have been performed at the ILL reactor in a well known neutron flux [4]. The sensitivity of the detectors has been measured and the dispersion of the results is evaluated to be less than 3%. Then, all the masses of the $^{235}$U deposits, which constitute one of the main sources of uncertainty, were checked by γ-spectroscopy and mass-spectrometry resulting in an absolute error less than 1%. The mechanical resistance of the actinide deposit in a high temperature environment was also tested by MEB analysis. Finally, we started a very deep study on the comprehension of all the physical processes involved in the functioning of such a detectors [6]. In particular, all the processes which can degrade the response of the detector, as space charge effects, have been modelled in detail, providing a simulation tool able to calculate the delivered current as a function of the applied voltage. All these developments allow limiting the global uncertainty on the neutron flux measurement to less than 3%.

## 3 Neutronic of the target

### 3.1 Preliminary physics results

The MEGAPIE has stand for four months under a proton beam power closed to 700 kW, instead of 1 MW, due to the difficulty to provide a stable proton beam over 1.2 mA. Nevertheless, the gain in terms of neutron flux was higher than 80% in comparison with a solid metal target, greatly exceeding expectations. The neutron detector has functioned reliably for all this time at a temperature around 400°C with frequent beam interruptions. During the whole irradiation phase the currents of the 8 fission chambers have been recorded every 2s. The beam current intensity was also recorded to study the neutron production of the target normalised to one incident proton, which is one of the fundamental items in the economy of a neutron source. The results we present here are preliminary, data analysis being still in progress.

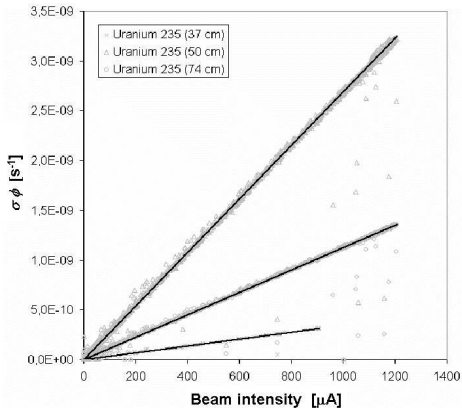

**Fig. 3.** Correlation between the fission rate and the proton beam intensity.

The current measured by fission chambers is proportional to the fission rate ($\sigma\phi$) which depends on the neutron flux and the effective fission cross section of the fissile isotope. The extraction of the neutron flux is then not straightforward and depends on a good characterisation of the neutron energy distribution which is calculated with simulation codes. However, if the epithermal/thermal ratio does not evolve over time or with the beam intensity, the fission rate is a good estimate of the relative variations of the neutron flux. The evolution of the fission rate as a function of the proton beam intensity is shown on fig. 3, where we see a good correlation for the three uranium chambers, as expected. This validates the correct functioning of the detectors. On fig. 4 the evolution of the fission rate normalised to the proton beam intensity is plotted, as a function of time, for the middle and the upper chambers. We can see a small decrease of the fission currents due to the burn-up of the uranium deposit, estimated around 6%.

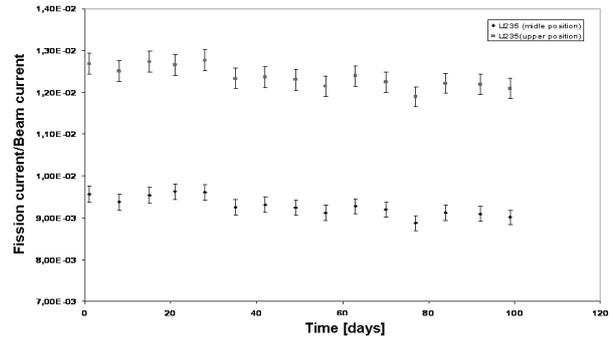

**Fig. 4.** Time evolution of the $^{235}$U fission chamber signal, normalised to the proton beam current.

Figure 5 shows the evolution of $^{241}$Am and $^{237}$Np currents as a function of time. We clearly see the increase of Am current due to the transmutation of $^{241}$Am into the long-lived fissionable isomeric state of $^{242}$Am. From these data it should be possible to extract a fission cross section for $^{242m}$Am into a moderated flux. The incineration of the $^{237}$Np is much more difficult to see due to the 2 days half-life fissile isotope $^{238}$Np which decays before absorbing neutrons to fission, thus leading to a constant current during all the irradiation.

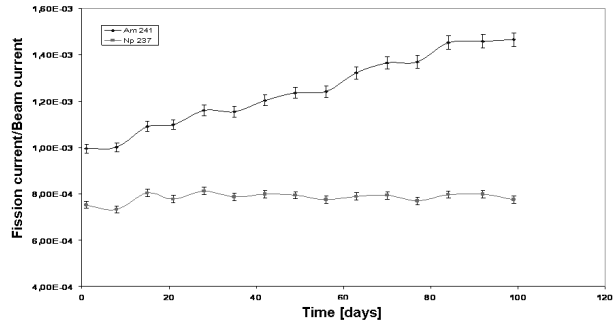

**Fig. 5.** Time evolution of the $^{241}$Am and $^{237}$Np fission chamber signal, normalised to the proton beam current.

### 3.2 Simulation studies

The MEGAPIE target is a complex system that has been simulated using Monte Carlo transport codes like FLUKA and MCNPX. A first set of simulations [2] was performed in the target R&D phase in order to define the key parameters of the experiment (neutron flux intensity, mass of the deposit in the fission chambers, thermal/fast neutrons, activation, etc…). The simulation work has been constantly improved, taking into account a much better description of the whole SINQ geometry, including the target head, the TKE environment and the neutron guides. In particular, we studied the influence of the neutron detector geometry, the composition of the LBE and the beam profile on the neutron flux. These simulations show that a better description of the neutron detector, which takes into account the presence of activation foils near the fission chambers, affects the neutron flux on the fission chambers not more than 1%.

The composition of the liquid Pb-Bi is a very important item because the presence of neutron absorbers (like boron, gadolinium or cadmium) in the LBE can modify significantly the energy distribution of the neutron flux including its absolute values. We present in table 1 the simulated fission rate for the four positions along the neutron detector and for different concentrations of boron in the LBE. We call "Std boron" the boron concentration which has been effectively measured from a sample of MEGAPIE LBE ($^{10}$B: 6.51 ppm; $^{11}$B: 27.2 ppm). This "real" composition is compared to the one without boron and two others with the "Std boron" concentration multiplied respectively by 10 and 100. The comparison shows that the presence of some ppm of boron in the LBE affects the neutron flux mostly in the thermal region, as expected. In the lowest position, where the Gd shielding cuts away the thermal part of the spectrum, the effect of B is around 3% and does not exceed 20% when the boron concentration reaches a few per mil. On the contrary, looking at the upper position, which is characterized by an almost fully moderated spectrum, the presence of some per mil of boron changes the fission rate by a factor of 3.

**Table 1.** Simulated fission rates (per incident proton and per second) for different boron concentrations in LBE. The spallation model used is the MCNPX default one.

| Position | 1 ($^{235}$U + Gd) | 3 ($^{235}$U) | 4 ($^{241}$Am) | 5 ($^{235}$U) |
| --- | --- | --- | --- | --- |
| No boron | 6.74 e-10 | 8.63 e-9 | 3.92 e-11 | 3.08 e-9 |
| Std boron | 6.63 e-10 | 8.48 e-9 | 3.99 e-11 | 3.07 e-9 |
| Std x 10 | 6.43 e-10 | 7.28 e-9 | 3.66 e-11 | 2.83 e-9 |
| Std x 100 | 5.44 e-10 | 2.93 e-9 | 1.63 e-11 | 1.02 e-9 |

Another parameter which can affect the neutron flux is the beam profile. The first simulations were performed taking a beam footprint extrapolated from old measurements on SINQ beam line. Simulations show an effect up to 14% (for the lower chamber) when one takes into account a more recent activity measurement on the SINQ solid target.

Taking into account all these improvements to the target description, we compared the measured fission rate to the simulated one. We remind that the fission rate is extracted from the measured fission current taking into account the chamber sensitivity previously measured at ILL. Concerning the simulation, the fission rate is calculated taking the neutron flux distribution given by the MCNPX transport code and the $^{235}$U fission cross section from ENDFB-VI library. In table 2 we compare the measured fission rate with two simulations, obtained using different physics model. We compared the so called Bertini-Dressner, which is the default model used in MCNPX, with the coupling of the intra-nuclear cascade model INCL4 [7] and the evaporation/fission model ABLA [8]. We can see that the difference between the models is not large, below 9%, and there is a systematic over-prediction of the measured values by a factor of 2-3, which cannot be explained for the moment. A possible explanation comes from the fact that our present simulation doesn't take into account the actual temperature of Pb-Bi and D$_2$O moderator, which could be expected to have a non negligible influence.

**Table 2.** Comparison between preliminary measured fission rates with two simulations made with Bertini-Dressner (BD) and INCL4-ABLA (IA) models.

| Position | 1 ($^{235}$U+Gd) | 3 ($^{235}$U) | 4 ($^{241}$Am) | 5 ($^{235}$U) |
| --- | --- | --- | --- | --- |
| Measured $\sigma\phi$ | 3.70E-10 | 2.85E-9 | 1.37E-11 | 1.19E-9 |
| (uncertainty) | (3%) | (3%) | (3%) | (3%) |
| Simulated $\sigma\phi$ (Bertini-Dressner) | 6.63E-10 | 8.48E-9 | 3.99E-11 | 3.07E-9 |
| Simulated $\sigma\phi$ (Incl4-Abla) | 7.01E-10 | 8.29E-9 | 3.83E-11 | 3.07E-9 |

## 4 Conclusions and perspectives

We presented a set of studies on the neutronics of the MEGAPIE spallation target, based on the measurement of the inner neutron flux and the simulation of the system. As a general comment, we want to point out that all the simulation studies show the importance of the implantation of the neutron detector inside the target to study macroscopic effects that could greatly modify estimated quantities as, for example, activation residues. The simulation work is still ongoing, focused on the study of the influence of the temperature of the LBE and of the different models used in MCNPX.

From the experimental point of view, the analysis of the fission chamber is progressing to get a final measurement of the inner flux as a function of the position along the beam axis. Moreover, to complete the spectral analysis of the flux in the target, activation monitors that were also irradiated with the fission chambers will be analysed by γ-spectroscopy during the Post Irradiation Experiment (PIE) which should take place by the end of 2008.

The authors are grateful to the MEGAPIE project management and to the PSI-SINQ personnel for cooperation and assistance during the detector installation and data taking. This work was partially supported by the MEGAPIE initiative and also by the GDR GEDEPEON (France).

## References

1. G.S. Bauer et al., Journal of Nuclear Mat. **296**, (2001) 17-35.
2. A. Letourneau et al., in *Summary Report for MEGAPIE R&D Task Group X9* (PSI Bericht, 05-12, ISSN 1019-06432005).
3. A. Letourneau et al, these proceeedings
4. S. Chabod et al., Nucl. Intr. And Meth. A **562**, (2006) 618-620.
5. S. Chabod et al., Nucl. Intr. And Meth. A **566**, (2006) 633.
6. S. Chabod, Ph.D. thesis, University of Paris XI, 2006.
7. A. Boudard et al., Phys. Rev. C **66**, (2002) 044615.
8. A. R. Junghans et al., Nucl. Phys. A **629**, (1998) 635.